\documentclass[twocolumn,aps,showpacs,amsmath,amssymb,floatfix,pre]{revtex4}
\usepackage{graphicx}% Include figure files 
\begin{document}

\title{Weight-driven growing networks}

\author{T. Antal}
\altaffiliation{On leave from Institute for Theoretical Physics, E\"otv\"os University, Budapest, Hungary}
\author{P. L. Krapivsky}
\email{paulk@bu.edu}
\affiliation{Center for Polymer Studies
and Department of Physics, Boston University, Boston, MA 02215, USA}
\begin{abstract}

  We study growing networks in which each link carries a certain weight
  (randomly assigned at birth and fixed thereafter). The weight of a node is
  defined as the sum of the weights of the links attached to the node, and
  the network grows via the simplest weight-driven rule: A newly-added node
  is connected to an already existing node with the probability which is
  proportional to the weight of that node.  We show that the node weight
  distribution $n(w)$ has a universal, that is independent on the link weight
  distribution, tail: $n(w)\sim w^{-3}$ as $w\to\infty$.  Results are
  particularly neat for the exponential link weight distribution when $n(w)$
  is algebraic over the entire weight range.

\end{abstract}

\maketitle

\section{Introduction} 
\label{sec:int}

The network structure lies beneath many physical, biological, social,
economical, and other real-world systems.  Examples include resister
networks, metabolic networks, communication networks like the Internet,
information networks like the world-wide web, transportation networks, food
webs, etc. Some networks are engineered while others like the world-wide web
are created in a chaotic manner, viz.\ by the uncoordinated actions of many
individuals, and yet they show a great deal of self-organization \cite{books}.  This
surprising order of seemingly disordered networks was noticed long ago in the
context of random graphs \cite{ER,Bol}, and it is also apparent in the models
of growing random networks that have been thoroughly studied in the past few
years (see reviews \cite{sum1,sum2,sum3} and references therein).

Networks or graphs are defined as sets of nodes joined by links.  A link in a
graph merely indicates that a given pair of nodes is connected.  If however
connections differ in ``strength'', one may formalize this property by
assigning a weight to each link.  Many real-world networks are intrinsically
weighted. In a collaboration network, the weight of a link between co-authors
measures the strength of the collaboration such as the number of
jointly-authored publications \cite{New,B}. In the airline transportation
network, the weight of a link between two airports gives the passenger
capacity on this route \cite{air,orsay}.  Thus mathematically the weighted
network is a graph in which each link carries a certain number which is
called the weight.  Weighted networks appear in literature under different
names. For instance, the multigraph --- that is the graph in which two nodes
can be joined by multiple links --- can be replaced by the graph in which a
link between two nodes carries the integer weight equal to the number of
links in the original multigraph joining those two nodes. 
Numerous engineering and mathematical studies of the flows in networks also
treat weighted networks --- the weight (usually termed ``capacity'')
represents the maximum allowed flow \cite{F1,F2}.  Resistors networks (see
e.g.  \cite{Doyle,R,RT}) form another very important class of weighted
networks.

The models of weighted networks are usually close to the unweighted ones as
far as the underlying graph structure is concerned. In other words, the link
weights are passive variables. This allows to study weighted networks using
the knowledge of unweighted networks as the starting point
\cite{New2,preprint}. In reality, the link weights can of course affect the
graph structure. The range of possible models in which the link weights are
active variables is extremely broad (see e.g. \cite{tu,NR,hui,macdonald,BBV})
while the mechanisms driving the evolution of the real-world networks are
still hardly known. In such situation one wants to study a minimal rather
than detailed models. Here we introduce and investigate a minimal model of
the weight-driven growth.

In Sec.~\ref{sec:mod}, we introduce the minimal model precisely, and 
determine the node weight distribution and the joint node weight-degree
distribution.  We then compute the in-component weight distribution.  In
Sec.~\ref{sec:gen}, we discuss some generalizations of the minimal model. We
conclude in Sec.~\ref{sec:concl}.

\section{The Model} 
\label{sec:mod}

The model is defined as follows. Each link carries a positive weight $w$
which is drawn from a certain distribution $\rho(w)$.  We shall assume that
the weights are positive \cite{negative}.  The weight is assigned to the link
when the link is created, and it remains fixed thereafter. The weight of a
node is the sum of the weights of the links attached to the node.  When a new
node is added, it is linked to a single ``target'' node with probability
proportional to the weight of the target node.

We assume that only one link emanates from each newly-introduced node, so the
resulting network is a tree; the general case when a few links emanate from
each node is discussed in Sec.~{\ref{sec:gen}}. The weight of a node (also
termed as the node strength by some authors) increases when a new link is
attached to it, yet, as the size of the network grows, the node weight
distribution approaches a {\em stationary} distribution. Our first goal is to
determine this distribution.

\subsection{Weight Distribution} 
\label{sub:wd}

Let $N$ be the total number of nodes in the network and $N_w(N)\,dw$ the
number of nodes whose weight lies in the range $(w,w+dw)$. When $N$ is large,
it can be treated as a continuous variable, and $N_w(N)$ satisfies
\begin{equation}
\label{Nw}
\frac{d N_w}{dN}=\frac{1}{W}\left[\int_0^w dx\,\rho(w-x)\,x\,N_x
-w\,N_w\right]+\rho(w) ~,
\end{equation}
where $W(N)=\int_0^\infty dw\,w\,N_w(N)$ is the total weight of all nodes
\cite{note}.  The term in Eq.~(\ref{Nw}) which is proportional to $xN_x/W$
accounts for nodes with weight $x$ which gain a link of weight $w-x$ thereby
creating nodes of weight $w$. The term $wN_w/W$ is the corresponding loss
term. The last term accounts for the newly introduced node which has the same
weight as its link.

$N_w(N)=Nn(w)$ for large $N$, i.e. the weight distribution approaches a
stationary ($N$ independent) distribution $n(w)$. Hence, Eq.~(\ref{Nw})
simplifies to
\begin{equation}
\label{nw}
(\lambda + w)\,n(w)=\int_0^w dx\,\rho(w-x)\,x\,n(x)
+\lambda\,\rho(w) ~,
\end{equation}
where $\lambda=\int_0^\infty dw\,w\,n(w)$ is the average node weight.  The
total weight of all nodes is twice the total weight of all links (since each
link connects two nodes) and the same obviously holds for the average
weights.  Therefore 
\begin{equation*}
\lambda=2\langle w\rangle=2\int_0^\infty dw\,w\,\rho(w) ~.
\end{equation*}

We now specialize Eqs.~(\ref{nw}) to the uniform link weight distribution
\begin{equation}
\label{unidis}
\rho(w)=
\begin{cases}
1             & w<1\cr
0             & w>1 ~.
\end{cases} 
\end{equation}
Using relation $\lambda=2\langle w\rangle=1$ and the shorthand notation
$F(w)=1+\int_0^w dx\,x\,n(x)$ we recast (\ref{nw}) into
\begin{equation}
\label{uniform}
(1 + w)\,n(w)=
\begin{cases}
F(w)         & w<1\cr
F(w)-F(w-1)  & w>1 ~.
\end{cases} 
\end{equation}

\begin{figure}[htb]
  \centering
    \includegraphics[width=1.0\linewidth]{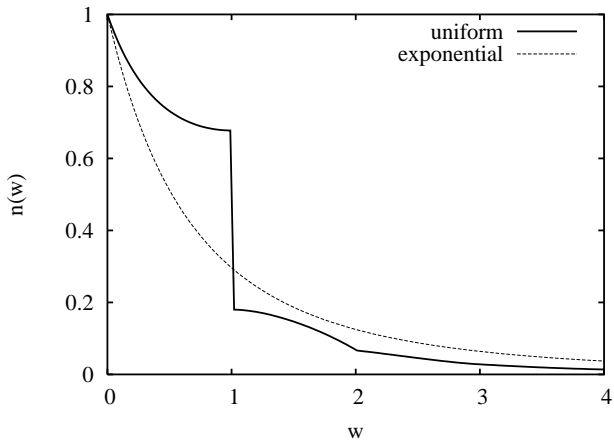} 
  \caption{The node weight distribution $n(w)$ emerging for 
    uniform and exponential link weight distributions.}
\label{flat+exponential}
\end{figure}

One can solve (\ref{uniform}) recursively. Using $F'(w)=w\,n(w)$ we find that
for $w<1$ the cumulative distribution satisfies $[w^{-1}+1]\,F'(w)=F(w)$
(here $F'=dF/dw$).  Solving this equation subject to the boundary condition
$F(0)=1$ yields $F(w)=(1+w)^{-1}\,e^w$. Therefore the weight distribution is
\begin{equation*}
n(w)=(1+w)^{-2}\,e^w\qquad {\rm for}\quad w<1 ~.  
\end{equation*}
On the next interval $1<w<2$, we ought to solve
\begin{equation*}
(w^{-1}+1)\,\frac{dF}{dw}=F - w^{-1}\,e^{w-1}    ~.
\end{equation*}
The solution should match the previous one; this gives the boundary condition
$F(1)=e/2$. The resulting cumulative distribution is
$F=e^{w-1}\,(e+1-w)/(1+w)$, from which we obtain the node weight
distribution
\begin{equation*}
n(w)=\frac{e^{w-1}}{1+w}\left(\frac{e+1-w}{1+w}-\frac{1}{w}\right)
\qquad {\rm for}\quad 1<w<2 ~.
\end{equation*}
Interestingly, the weight distribution loses continuity at the
cutoff value $w=1$ of the uniform link weight distribution: $n(1-0)=e/4$
while $n(1+0)=(e-2)/4$. Proceeding, one finds the node weight distribution on
the interval $2<w<3$, etc. The distribution is analytic everywhere apart from
the integer values; at the integer value $w=k\geq 2$, the node weight
distribution is continuously differentiable $k-2$ times (see
Fig.~\ref{flat+exponential}). The analytic expressions for the node weight
distribution are very cumbersome for large $w$. Fortunately, the asymptotic
behavior in the tail region is very simple.  To extract the asymptotic we
expand the right-hand side of (\ref{uniform}) using the Taylor series:
\begin{eqnarray*}
F(w)-F(w-1)&=&F'(w)-\frac{1}{2}\,F''(w)+\ldots\\      
&=&wn(w)-\frac{1}{2}\,[wn(w)]'+\ldots  ~.    
\end{eqnarray*}
Plugging this expansion into (\ref{uniform}) we obtain
\begin{equation}
\label{tailde}
w\,\frac{dn}{d w}+3\,n=0   ~,
\end{equation}
which is solved to give 
\begin{equation}
\label{tail} 
n(w)\to \frac{A}{w^3}\qquad {\rm when}\quad w\to\infty   ~.
\end{equation}
The amplitude $A$ cannot be found by solving the linear equation
(\ref{tailde}); its determination requires analysis of the full problem
(\ref{uniform}).

Note also that for the deterministic link weight distribution
$\rho(w)=\delta(w-1)$ the weight of the node is equal to the degree, so the
model becomes equivalent to the basic growing network model with preferential
attachment that also leads to a node weight distribution with $w^{-3}$
tail.  In general, the large weight tail of the node weight distribution is
always given by Eq.~(\ref{tail}) if the link weight distribution has a
cutoff.  To prove this assertion suppose that $\rho(w)=0$ for $w>1$. For
$w>1$, Eq.~(\ref{nw}) becomes
\begin{equation}
\label{gende}
(\lambda+w)\,n(w)=\Phi(w)
\end{equation}
with $\Phi(w) = \int_0^1 dx\, \rho(x)\,(w-x)\,n(w-x)$; for the uniform
distribution $\Phi(w) = F(w)-F(w-1)$ and Eq.~(\ref{gende}) turns into
Eq.~(\ref{uniform}).  Expanding the integrand in $\Phi(w)$
\begin{equation}
\label{expansion} 
(w-x)\,n(w-x)=w\,n(w)-x\,[wn(w)]'+\ldots 
\end{equation}
yields
\begin{equation}
\label{Phi} 
\Phi(w)=wn(w)-\frac{\lambda}{2}\,[wn(w)]'+\ldots ~,
\end{equation}
where relations $\int dw\,\rho(w)=1$ and $\int dw\,w\,\rho(w)=\lambda/2$ were
used.  In conjunction with (\ref{Phi}), Eq.~(\ref{gende}) reduces to
Eq.~(\ref{tailde}) leading to the tail given by (\ref{tail}). This
derivation, as well as the earlier one that led to Eq.~(\ref{tailde}),
ignores the higher terms in the expansion (\ref{expansion}).  Keeping such
terms one would obtain
\begin{equation}
\label{gendndw}
w\,\frac{dn}{d w}+3\,n=\kappa_2\, [wn(w)]'' 
-\frac{1}{3}\,\kappa_3\, [wn(w)]'''+\ldots 
\end{equation}
with $\kappa_2=\lambda^{-1}\int dw\,w^2\,\rho(w), \kappa_3=\lambda^{-1}\int
dw\,w^3\,\rho(w)$, etc.  Solving (\ref{gendndw}) gives higher order
corrections to the tail asymptotics
\begin{equation}
\label{nwcutoff}
n=\frac{A}{w^3}\left[1-\frac{6\,\kappa_2}{w}
+\frac{36\,(\kappa_2)^2-4\kappa_3}{w^2}+\ldots\right]   
\end{equation}
but it does not affect the leading $w^{-3}$ tail. 

The exponential link weight distribution
\begin{equation}
\label{exp} 
\rho(w)=e^{-w}
\end{equation}
is particularly appealing as the node weight distribution in this case is
remarkably simple.  The governing equations (\ref{nw}) become
\begin{equation}
\label{expint}
(2 + w)\,e^w\,n(w)=\int_0^w dx\,x\,e^x\,n(x)+2 ~.
\end{equation}
Writing $G(w)=2+\int_0^w dx\,x\,e^x\,n(x)$ we recast integral equation (\ref{expint})
into a simple differential equation 
\begin{equation}
\left(\frac{2}{w}+1\right)\frac{dG}{dw}=G   ~.
\end{equation}
We find $G(w)=8(w+2)^{-2}\,e^w$, from which
\begin{equation}
\label{neat}
n(w)=\frac{1}{\left(1+\frac{w}{2}\right)^3}  ~.
\end{equation}
Thus for the exponential link weight distribution, the emerging node weight
distribution is scale-free (that is, purely algebraic) over the entire weight
range.

We now outline the behavior for link distributions with heavy tails. Consider
particularly distributions with a power-law tail $\rho(w) \sim w^{-\nu}$; the
exponent must obey the inequality $\nu>1$ to ensure normalization $\int
dw\,\rho(w)=1$. It is easy to check that for $\nu>3$ the leading asymptotic
of the node weight distribution is $n(w)\sim w^{-3}$; moreover, expansion
(\ref{nwcutoff}) holds up to the order of $n$, where $n$ is the largest
integer smaller than $\nu$ (i.e. $n<\nu<n+1$) \cite{integer}.  For $2<\nu<3$
the leading term shows even slower decay $n(w)\sim (3-\nu)^{-1} w^{-\nu}$.

The situation is very different for $1<\nu\le2$ when the total weight grows
with size faster than linearly: $W\sim N^{1/(\nu-1)}$. Using the usual
definition $N_w=Nn(w)$, we observe that the
square bracket term in Eq.~(\ref{Nw}) is negligible for large $N$ values. Hence
$n(w)\to\rho(w)$, suggesting that the number of nodes with more than one links
grows slower than linearly in $N$, that is only the dangling nodes give
contribution to $n(w)$.

\subsection{Weight-Degree Distribution} 
\label{sub:wdd}

The node weight distribution is perhaps the most natural, and therefore
readily tractable, local characteristic for networks with weight-driven
growth. Of course, the degree distribution remains geometrically the simplest
local characteristic, yet for the weighted networks it is generally impossible
to write down a closed equation for the degree distribution $n_k$. To compute
$n_k$, one must determine the joint weight-degree distribution $n_k(w)$; the
degree distribution is then found by integration
\begin{equation}
n_k=\int_0^\infty dw\,n_k(w)  ~.
\end{equation}
The weight-degree distribution obeys a set of equations similar to
(\ref{nw}). The density of dangling nodes is given by
\begin{equation}
\label{nw1}
(\lambda + w)\,n_1(w)=\lambda\,\rho(w)
\end{equation}
while for $k\geq 2$ the weight-degree distribution satisfies
\begin{equation}
\label{nwk}
(\lambda + w)\,n_k(w)=\int_0^w dx\,\rho(w-x)\,x\,n_{k-1}(x)  ~.
\end{equation}
One can treat Eqs.~(\ref{nw1})--(\ref{nwk}) recursively, yet even for the
simplest link weight distributions like the uniform distribution
(\ref{unidis}) the exact expressions for $n_k(w)$ become very unwieldy as
the degree grows.  Exceptionally neat results emerge again for the
exponential link weight distribution (\ref{exp}).  In this case
\begin{equation}
(2 + w)\,e^w\,n_k(w)=\int_0^w dx\,e^x\,x\,n_{k-1}(x) ~.
\end{equation}
Starting with $n_1(w)=\frac{2}{2+w}\,e^{-w}$ we explicitly computed a few
more $n_k(w)$ which led us to the hypothetical solution (we use the
shorthand notation  $\Omega=w-2\ln\left(1+\frac{w}{2}\right)$)
\begin{equation}
\label{nwk-sol}
n_k(w)=\frac{1}{\left(1+\frac{w}{2}\right)^3}\,\,\frac{\Omega^{k-1}\,
e^{-\Omega}}{(k-1)!}  ~.
\end{equation}
(Of course, the above ansatz agrees with the sum rule $n(w)=\sum_{k\geq 1}
n_k(w)$.)  Having guessed the solution, it is then straightforward to verify
its validity.  Note that asymptotically ($k\gg 1$), the joint weight-degree
distribution approaches a Gaussian centered around $\Omega=k-4$, or
$w_k\approx k+2\ln k$, with width $\propto \sqrt{k}$.  Interestingly, the
average weight of the nodes of degree $k$, viz.\  $w_k\approx k+2\ln k$,
slightly exceeds \cite{comment} the expected value $w_k=k$.

The degree distribution does not admit a simple closed form even for
the exponential link weight distribution. The exact expression 
\begin{equation}
\label{nk-sol}
n_k=\int_0^\infty \frac{d\Omega}{\frac{w}{2}\left(1+\frac{w}{2}\right)^2}
\,\,\frac{\Omega^{k-1}\,e^{-\Omega}}{(k-1)!}
\end{equation}
simplifies for $k\gg 1$. Indeed, utilizing two properties of the quantity
$\Omega^{k}\,e^{-\Omega}/k!$\,, viz.\ (i) it has a sharp maximum at
$\Omega=k$ and (ii) $\int_0^\infty
d\Omega\,\,\Omega^{k}\,e^{-\Omega}/k!=1$, we estimate the integral in
(\ref{nk-sol}) as the value of the slowly varying part of the integrand
$(\frac{w}{2})^{-1}(1+\frac{w}{2})^{-2}$ near the maximum,
i.e. at $w_k\approx k+2\ln k$. Thus
\begin{equation}
\label{nk-asymp}
n_k=\frac{8}{k^3}-48\,\frac{\ln k}{k^4}+{\cal O}(k^{-4}) ~.
\end{equation}

\subsection{In-Component Weight Distribution} 
\label{sub:icwd}

\begin{figure}[htb]
  \centering
    \includegraphics[width=0.9\linewidth]{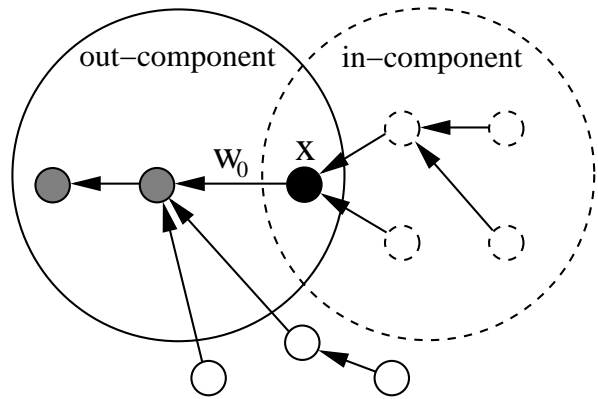} 
  \caption{In and out components of node ${\bf x}$. In this example, the 
    out-component has size 3 and the in-component has size 5 (node ${\bf x}$ itself
    belongs to its in and out components).}
\label{in-out}
\end{figure}

The emerging network has the natural structure of the {\em directed} graph
since each new link starts at the new node and ends up at some previous node.
Taking into account the orientation of each link allows to define an {\em
  in-component} and an {\em out-component} with respect to each node. For
instance, the in-component of node ${\bf x}$ is the set of all nodes from
which node ${\bf x}$ can be reached following a path of directed links
(Fig.~\ref{in-out}).  The computation of in- and out-component size
distributions is quite complicated as one must first determine joint
distributions that involve both size and weight.  In contrast, in- and
out-component weight distributions can be determined directly. Here we
compute the in-component weight distribution.

Let $w_0$ be the weight of the link emanating from node ${\bf x}$ and 
$w$ the total weight of all other links in the in-component of node ${\bf
  x}$; then the total weight of that in-component, that is the sum of weights
of all nodes in in-component of node ${\bf x}$, is $w_0+2w$. Denote by
$i(w_0,w)$ the density of such in-components. Here we tacitly assume that
$w>0$, that is the size of the in-components is larger than one.  
The density $i(w_0)$ of in-components of size one and weight $w_0$ is found
by noting that such in-components are just dangling nodes, so 
\begin{equation}
\label{i1-sol}
i(w_0)\equiv n_1(w_0)=\frac{\lambda}{\lambda + w_0}\,\rho(w_0) ~.
\end{equation}
The density $i(w_0,w)$ satisfies
\begin{eqnarray*}
\lambda\,i(w_0,w)&=&\int_0^w dx\,\rho(w-x)\,(w_0+2x)\,i(w_0,x)\\
&-&(w_0+2w)\,i(w_0,w)+i(w_0)\,w_0\,\rho(w) ~.
\end{eqnarray*}

For the exponential link weight distribution (\ref{exp}), this equation can
be re-written in a simple form
\begin{equation}
\label{Gww} 
\left(\frac{2}{w_0+2w}+1\right)\frac{\partial G}{\partial w}=G   ~,
\end{equation}
where $G=G(w_0,w)$ is the auxiliary variable 
\begin{equation}
\label{Gww-def} 
G=\int_0^w dx\,e^x\,(w_0+2x)\,i(w_0,x)+\frac{2\,w_0}{2+w_0}\,e^{-w_0}  ~.
\end{equation}
Solving (\ref{Gww}) we obtain
$G(w_0,w)=G(w_0,0)\,\frac{2+w_0}{2+w_0+2w}\,e^w$.  Equation (\ref{Gww-def})
gives the initial condition, so the auxiliary variable reads
$G=\frac{2\,w_0}{2+w_0+2w}\,e^{w-w_0}$. This result leads to the in-component
density
\begin{equation}
\label{iw-sol} 
i(w_0,w)=\frac{2\,w_0}{(2+w_0+2w)^2}\,e^{-w_0}  ~.
\end{equation}
Now the total weight of the in-component is $s=w_0+2w$, and the respective weight
distribution $I(s)$ is 
\begin{equation}
\label{is-def} 
I(s)=i(s)+\frac{1}{2}\int_0^s dw_0\,i[w_0,(s-w_0)/2] ~.
\end{equation}
Plugging (\ref{i1-sol}) and (\ref{iw-sol}) into (\ref{is-def}) we arrive at
\begin{equation}
\label{is-sol} 
I(s)=\frac{1}{(2+s)^2}+\left[\frac{1}{2+s}+\frac{1}{(2+s)^2}\right]e^{-s}  ~,
\end{equation}
Therefore up to an exponentially small correction, the in-component weight
distribution is algebraic with exponent 2 (the same exponent characterizes
the in-component size distribution \cite{KR}, which is not so surprising as
this exponent is found to be very robust).

\section{Generalizations} 
\label{sec:gen}

The model of the previous section always results in a tree structure by
construction.  In this section we consider two generalizations of that model
which allows the formation of loops in the evolving network.

The simplest generalization of the minimal model leading to a network with
many loops is to connect a newly created node to $m$ target nodes.  The
attachment probability is still proportional to the weight of the target node
as before, and the weights of the $m$ new links are chosen independently from
the link weight distribution $\rho(w)$. The weight of the newly introduced
node becomes the sum of the $m$ independent link weights, which is then
distributed according to $\rho_m(w)$, the $m$-fold convolution of $\rho(w)$.
The governing equation thus remains similar to Eq.~(\ref{Nw}) except that the
last term $\rho(w)$ becomes $\rho_m(w)$, and the square bracketed term gains
a factor $m$ due to the $m$ attached links. As the average node weight is
also $m$ times larger than previously, that is $\lambda=2m\,\langle w
\rangle$, and the node density satisfies equation
\begin{equation*}
%\label{nw-m}
(2\langle w \rangle + w)\,n(w)=\int_0^w dx\,\rho(w-x)\,x\,n(x)
+2 \langle w \rangle \rho_m(w)
\end{equation*}
which is almost identical to Eq.~(\ref{nw}), the only difference is the
second term on the right-hand side which now contains $\rho_m(w)$ instead of
$\rho(w)$.  The modified model leads to the same $\sim w^{-3}$ tail
for the node weight distribution for any $\rho(w)$ with a cutoff since then
$\rho_m(w)$ also has a cutoff and the same argument applies as before. For
the exponential link weight distribution $\rho(w)=e^{-w}$ the convoluted
distribution is $\rho_m(w)=e^{-w}\,w^{m-1}/(m-1)!$, and the resulting node
weight distribution is algebraic (up to an exponentially small correction)
\begin{equation*}
n(w) \to \frac{2m(m+3)}{(2+w)^3}\qquad{\rm for}\quad w\gg 1.
\end{equation*}

A more substantial generalization of the minimal model is to allow the
creation of links between already existing nodes \cite{KRR}. We again choose the simplest
weight-driven rule, namely we assume that a new link is created with rate
proportional to the product of weights of the originating and target nodes.
Let $r$ be the average number of link creations between already existing
nodes per node creation, so in the network with $N$ nodes the average number
of links is $(1+r)\,N$.  This can be modeled by two independent processes:
with probability $1/(1+r)$, a new node is attached to the network, and with
probability $r/(1+r)$ a new link is added between two already existing nodes.
The weight distribution $N_w(N)$ satisfies an equation similar to (\ref{Nw}),
with the term in the square brackets multiplied by $(1+2r)/W$. The governing
equation for the normalized weight distribution $n(w)$ is therefore almost
identical to Eq.~(\ref{nw}), the only change is the replacement $\lambda\to
\Lambda=\lambda/(1+2r)$:
\begin{equation}
\label{nwr}
(\Lambda + w)\,n(w)=\int_0^w dx\,\rho(w-x)\,x\,n(x)
+\Lambda\,\rho(w) ~.
\end{equation}
The average node weight $\lambda=\int_0^\infty dw\,w\,n(w)$ is now equal to
$2(1+r)\langle w\rangle$ since the number of links now $1+r$ times exceeds
the number of nodes and each link contributes twice. Multiplying (\ref{nwr})
by $w$ and integrating over $w$ we indeed obtain $\lambda=2(1+r)\int_0^\infty
dw\,w\,\rho(w)$ thereby providing a useful check of self-consistency.

For the exponential link weight distribution (\ref{exp}) the node weight
distribution satisfies 
\begin{equation*}
\left(1 + \frac{1+2r}{2+2r}\,w\right)\,e^w\,n(w)
=1+\frac{1+2r}{2+2r}\int_0^w dx\,x\,e^x\,n(x) ~.
\end{equation*}
Solving this equation we again obtain the purely algebraic node weight
distribution:
\begin{equation*}
n(w)=\left(1+\frac{1+2r}{2+2r}\,w\right)^{-\nu}\,,\qquad \nu=\frac{3+4r}{1+2r}  ~.
\end{equation*}
The exponent $\nu$ monotonously decreases from 3 to 2 as $r$ increases from 0
to $\infty$. 

\section{conclusions}
\label{sec:concl}

We examined a minimal model for the weight-driven network growth. 
The virtue of the minimal model is that its many
features like the node weight distribution, the joint
node weight-degree distribution, the in-component weight distribution, etc.\
are tractable analytically. In particular, we showed that the node weight
distribution exhibits a universal $w^{-3}$ tail independently on the link
weight distribution (as long as the tail of the latter is sharper than
$w^{-3}$), and the in-component weight distribution displays a robust $s^{-2}$
tail. 

Remarkably simple behaviors characterize the exponential link weight
distribution: The emerging node weight distribution is purely algebraic, the
joint node weight-degree distribution and the in-component weight
distribution are also given by neat closed formulas.

We also studied generalizations of the minimal model for the weight-driven
network growth.  When a new node is connected to several target nodes and/or
links are additionally created between already existing nodes, the network
acquires loops yet it remains tractable. One generic feature of this class of
models is that the node weight distribution remains algebraic, $n(w)\sim
w^{-\nu}$, with the exponent $\nu$ varying from 3 to 2 as the
average node degree increases from one to $\infty$.

\section{Acknowledgment}
TA thanks the Swiss NSF for financial support under the fellowship 8220-067591.


\begin{thebibliography}{99}
\bibitem{negative}    
    Negative weights are occasionally appropriate,
    e.g.  they can represent animosity between individuals in a social network.
\bibitem{note} 
    $N$ is a discrete variable and $N_w(N)$ are random
    variables. Treating $N$ as a continuous variable and $N_w(N)$ as the
    average values of the corresponding random variables is asymptotically
    exact when the weight is sufficiently small; see e.g.  P.~L.~Krapivsky
    and S.~Redner, J.\ Phys.\ A {\bf 35}, 9517 (2002) for the detailed
    analysis of these issues in the model where growth is governed by
    preferential attachment.
\bibitem{integer} 
    For integer $\nu\geq 4$, the $\nu^{\rm th}$ term in
    expansion (\ref{nwcutoff}) acquires a logarithmic correction; for
    $\nu=3$, even the leading order term has a logarithmic correction $n(w)
    \sim w^{-3}\ln(w)$.
\bibitem{comment} 
    The expected value for the sum of $k$ independent
    identically distributed random variables taken from the exponential
    distribution is $w=k$; in the present case the average weight of the node
    of large degree is (slightly) higher since the growth is weight-driven.
\end{thebibliography}
\end{document}